\documentclass[aps,prl,preprint,superscriptaddress,showpacs,byrevtex]{revtex4}
\usepackage{graphicx} 
\usepackage{amsmath, amsthm, amssymb}
\usepackage{epsfig}

\preprint{\vbox{ \hbox{}
                                                \hbox{Belle Preprint 2007-11}
                                                \hbox{KEK Preprint 2006-75}
}}

\begin{document}

\title{ \quad\\[0.5cm] Evidence for  $\mathbf{CP}$ Violation 
in $\mathbf{B^0 \to D^+ D^-}$ Decays}

\affiliation{Budker Institute of Nuclear Physics, Novosibirsk}
\affiliation{Chiba University, Chiba}
\affiliation{Chonnam National University, Kwangju}
\affiliation{University of Cincinnati, Cincinnati, Ohio 45221}
\affiliation{Department of Physics, Fu Jen Catholic University, Taipei}
\affiliation{The Graduate University for Advanced Studies, Hayama}
\affiliation{University of Hawaii, Honolulu, Hawaii 96822}
\affiliation{High Energy Accelerator Research Organization (KEK), Tsukuba}
\affiliation{Institute of High Energy Physics, Chinese Academy of Sciences, Beijing}
\affiliation{Institute of High Energy Physics, Vienna}
\affiliation{Institute of High Energy Physics, Protvino}
\affiliation{Institute for Theoretical and Experimental Physics, Moscow}
\affiliation{J. Stefan Institute, Ljubljana}
\affiliation{Kanagawa University, Yokohama}
\affiliation{Korea University, Seoul}
\affiliation{Kyungpook National University, Taegu}
\affiliation{Swiss Federal Institute of Technology of Lausanne, EPFL, Lausanne}
\affiliation{University of Ljubljana, Ljubljana}
\affiliation{University of Maribor, Maribor}
\affiliation{University of Melbourne, Victoria}
\affiliation{Nagoya University, Nagoya}
\affiliation{Nara Women's University, Nara}
\affiliation{National Central University, Chung-li}
\affiliation{National United University, Miao Li}
\affiliation{Department of Physics, National Taiwan University, Taipei}
\affiliation{H. Niewodniczanski Institute of Nuclear Physics, Krakow}
\affiliation{Nippon Dental University, Niigata}
\affiliation{Niigata University, Niigata}
\affiliation{University of Nova Gorica, Nova Gorica}
\affiliation{Osaka City University, Osaka}
\affiliation{Osaka University, Osaka}
\affiliation{Panjab University, Chandigarh}
\affiliation{Peking University, Beijing}
\affiliation{RIKEN BNL Research Center, Upton, New York 11973}
\affiliation{University of Science and Technology of China, Hefei}
\affiliation{Seoul National University, Seoul}
\affiliation{Shinshu University, Nagano}
\affiliation{Sungkyunkwan University, Suwon}
\affiliation{University of Sydney, Sydney, New South Wales}
\affiliation{Toho University, Funabashi}
\affiliation{Tohoku Gakuin University, Tagajo}
\affiliation{Tohoku University, Sendai}
\affiliation{Department of Physics, University of Tokyo, Tokyo}
\affiliation{Tokyo Institute of Technology, Tokyo}
\affiliation{Tokyo Metropolitan University, Tokyo}
\affiliation{Virginia Polytechnic Institute and State University, Blacksburg, Virginia 24061}
\affiliation{Yonsei University, Seoul}
  \author{S.~Fratina}\affiliation{J. Stefan Institute, Ljubljana} 
  \author{K.~Abe}\affiliation{High Energy Accelerator Research Organization (KEK), Tsukuba} 
  \author{K.~Abe}\affiliation{Tohoku Gakuin University, Tagajo} 
  \author{I.~Adachi}\affiliation{High Energy Accelerator Research Organization (KEK), Tsukuba} 
  \author{H.~Aihara}\affiliation{Department of Physics, University of Tokyo, Tokyo} 
  \author{D.~Anipko}\affiliation{Budker Institute of Nuclear Physics, Novosibirsk} 
 \author{K.~Arinstein}\affiliation{Budker Institute of Nuclear Physics, Novosibirsk} 
  \author{T.~Aushev}\affiliation{Swiss Federal Institute of Technology of Lausanne, EPFL, Lausanne}\affiliation{Institute for Theoretical and Experimental Physics, Moscow} 
 \author{A.~M.~Bakich}\affiliation{University of Sydney, Sydney, New South Wales} 
  \author{E.~Barberio}\affiliation{University of Melbourne, Victoria} 
% ...
\collaboration{The Belle Collaboration}

\date{\today}

\begin{abstract}
We report measurements of the branching fraction and $CP$ violation
parameters in $B^0 \to D^+ D^-$ decays.   
The results are based on 
a  data sample that contains $535  \times 10^6$  
 $B\overline{B}$ pairs  collected 
at the $\Upsilon(4S)$ resonance,
with the Belle detector at the KEKB asymmetric-energy $e^+ e^-$   
collider.
We obtain $ \left[1.97 \pm 0.20 \,\textrm{(stat)}\, \pm
    0.20\,\textrm{(syst)}\,\right] \,\times 10^{-4} $ for the branching fraction of 
$B^0 \to D^+D^-$.
The measured values of the  $CP$ violation parameters are:
  \mbox{$\mathcal S =  -1.13 \, \pm 0.37 \,\pm 0.09$},  \mbox{$\mathcal A
  = 0.91 \, \pm 0.23 \,\pm 0.06$}, where the first error  is 
statistical and the second is systematic.
We  find evidence of $CP$ violation in  $B^0 \to D^+ D^-$   at the 
$4.1\,\sigma$ confidence level. While the value of $\mathcal S$ 
is consistent with expectations from other measurements, the value of the parameter 
$\mathcal A$ favors large direct $CP$ violation at the $3.2\,\sigma$ confidence level,
in contradiction to Standard Model expectations.

\end{abstract}


\maketitle

{\renewcommand{\thefootnote}{\fnsymbol{footnote}}}
\setcounter{footnote}{0}

\begin{figure}[!h]
\centerline{
\begin{tabular}{ccc}
\epsfig{file=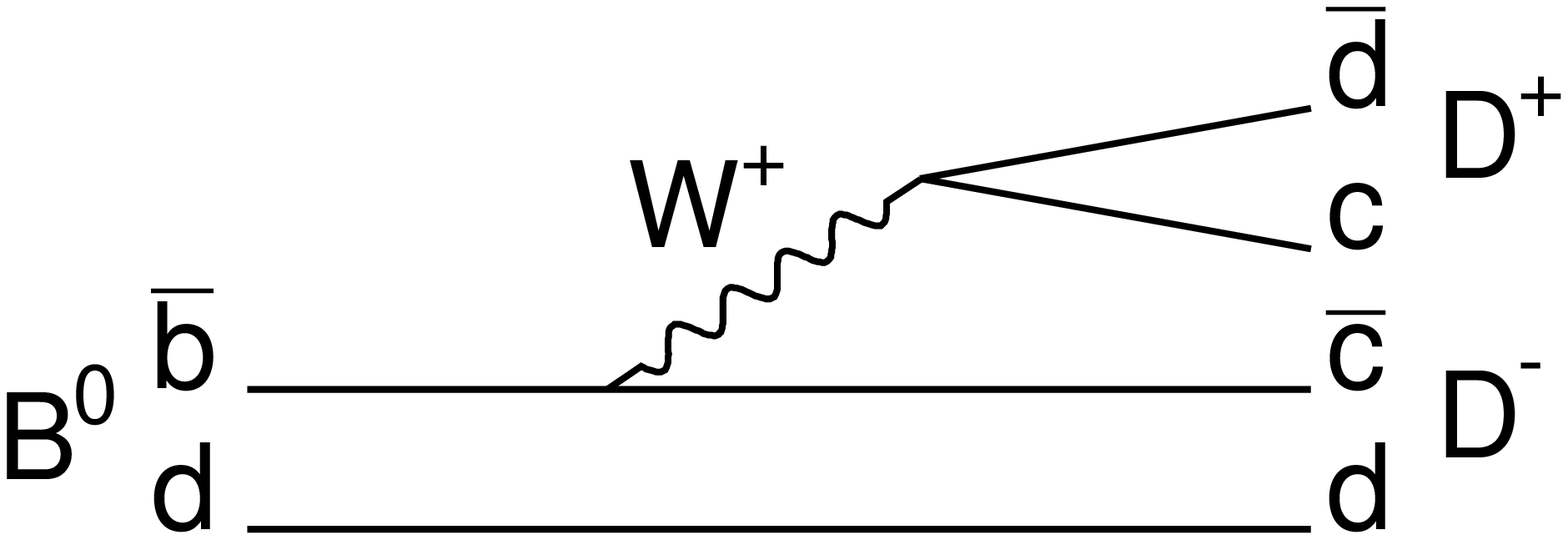,width=0.35\linewidth,clip=,}  
& \hspace{0.3cm} &
\epsfig{file=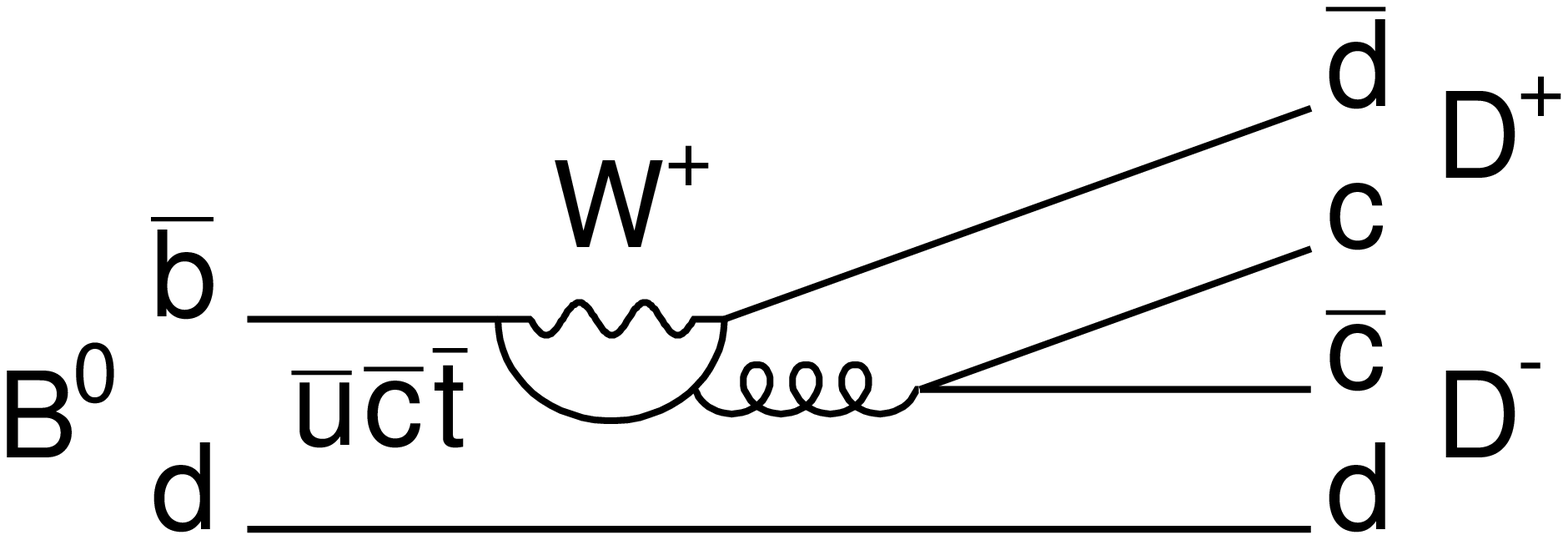,width=0.35\linewidth,clip=,}\\
(a) & & (b)
\end{tabular}
}
\caption{The tree (a) and the penguin (b) contributions to the $B^0 \to D^+ D^-$ decay.  } 
\label{fig:feynman}
\end{figure}

Within the Standard Model (SM), $CP$ violation ($CPV$) arises 
from a complex phase in the Cabibbo-Kobayashi-Maskawa (CKM)  
 quark mixing matrix $\mathbf V$ \cite{KM}. 
The dominant contribution to $B^0 \to D^+ D^-$ 
decays is the tree-level $b \to c\bar{c} d$ transition 
shown in Fig.~\ref{fig:feynman}(a). 
If this diagram is the only contribution, then the mixing-induced 
$CPV$ parameter for $B^0 \to D^+ D^-$ is  $- \sin\,2\phi _1$, where 
$\phi _1 = \mathrm{arg}[-V_{cd}V_{cb}^*/V_{td}V_{tb}^*]$, 
while the direct $CPV$ term  $\mathcal A$ is zero. 
The second-order gluonic penguin
contribution, shown in Fig.~\ref{fig:feynman}(b), is
expected to change the value of the parameter $\mathcal S$ by less than a 
few percent and increase the value of $\mathcal A$ to about $3\%$ \cite{ZZXing:theory}.
However, particles from physics beyond the SM may give additional
contributions  within the loop diagrams mediating flavor-changing $b \to
d$ transitions. Such contributions may potentially induce large
deviations from the SM expectation for time-dependent $CP$ asymmetries.
As $\sin\,2\phi _1$ has already been determined with high precision 
by measurements in $b \to c\bar{c} s$ charmonium modes \cite{Belle-phi1, Babar-phi1}, the 
objective here is to focus on deviations from expectations in $b \to c\bar{c} d$
transitions.  Similar studies have been carried out for $B^0 \to D^{*\pm} D^{(*)\mp}$ 
decays, which involve the same quark level weak decay  \cite{DstarD, DstarDstar, babar, babarDstarDstar}.

The $CPV$ parameters $\mathcal S$ and $\mathcal A$ can be measured from the $\Delta t$
distribution of  $B^0 \to D^+ D^-$ decays, 
\begin{equation}
\mathcal P_{\rm sig} =
 \frac{e^{-|\Delta
  t|/\tau}}{4\tau} \{1 + q [ \mathcal S \sin (\Delta m \Delta t) + \mathcal A  \cos (\Delta m \Delta t)]\}\,\mathrm{,}\label{eq:dP/dt}
\end{equation}
where $\Delta t = t_{CP} - t_\mathrm{tag}$ is the time difference
between decays of the two $B$ mesons arising from the $\Upsilon(4S)$.
The parameters $t_{CP}$ and $t_\mathrm{tag}$ are the proper decay times of 
the corresponding  $B$ mesons, 
$\tau$  is the $B^0$ meson lifetime 
and $\Delta m$  is the mass difference of the two 
$B$ mass eigenstates \cite{PDG}.
The flavor  $q$   
 is   determined from the final state  of the tagging $B$
 meson: $q=+1\,(-1)$ for $B_{\mathrm{tag}} = B^0\,(\overline{B}{}^0)$. 

The results presented here are
based on a data sample that
contains  $( 535 \pm 7 ) \times 10^6$  $B\overline{B}$ pairs, 
collected  with the Belle detector at the KEKB asymmetric-energy
$e^+e^-$ (3.5 on 8~GeV) collider~\cite{KEKB}.
KEKB operates at the $\Upsilon(4S)$ resonance 
($\sqrt{s}=10.58$~GeV) with a peak luminosity that exceeds
$1.7\times 10^{34}~{\rm cm}^{-2}{\rm s}^{-1}$.
At KEKB, the $\Upsilon(4S)$ is produced
with a Lorentz boost of $\beta\gamma=0.425$ nearly along
the electron beam line ($-z$ direction).
Since the $B^0$ and $\overline{B}{}^0$ mesons are approximately at 
rest in the $\Upsilon(4S)$ center-of-mass (CM) system,
$\Delta t$ can be determined from the displacement in $z$ 
between the $B_{CP}$ and $B_{\rm tag}$ decay vertices:
$\Delta t \simeq (z_{CP} - z_{\rm tag})/\beta\gamma c
 \equiv \Delta z/\beta\gamma c$.

The Belle detector   \cite{Belle}
is a large-solid-angle magnetic spectrometer that
consists of a silicon vertex detector,  
a 50-layer central drift chamber (CDC), an array of
aerogel threshold \v{C}erenkov counters (ACC), 
a barrel-like arrangement of time-of-flight
scintillation counters (TOF), and an electromagnetic calorimeter
comprised of CsI(Tl) crystals  located inside 
a superconducting solenoid coil that provides a 1.5~T
magnetic field.  An iron flux-return located outside of
the coil is instrumented to detect $K_L^0$ mesons and to identify
muons.  
Two inner detector configurations were used; a 2.0 cm radius beam pipe
and a 3-layer silicon vertex detector was used for  the first 
$152 \times 10^6$ $ B\overline{B}$ pairs and  a 1.5 cm beam pipe, a 4-layer
silicon detector and a small-cell inner drift chamber were employed  for 
the remaining $383 \times 10^6$ $B\overline{B}$ pairs \cite{Ushiroda}.  

$D$ mesons are reconstructed using the $D^+
\to  K^-\pi^+\pi^+$ and $D^+\to K_S\pi^+$  decay modes \cite{CC}.  
In this paper, the shorter notation $K\pi\pi$  is used when
both $D$ mesons are 
reconstructed in the $K\pi\pi$  channel while $K_S\pi$
is used when at least one of the $D$ mesons is
reconstructed in the $K_S\pi$ channel.
Charged tracks  that  are not positively identified as electrons \cite{elid}
and satisfy a loose requirement on the impact parameter relative to the
interaction point (IP) are considered as pion and kaon candidates. 
For charged particle identification (PID), we combine information
from the CDC, TOF and ACC counters into a likelihood ratio
${\cal{L}}(K^\pm)/[{\cal{L}}(K^\pm)+{\cal{L}}(\pi^\pm)]$, which is
required to be greater than $0.55$ for kaon and less than $0.9$ for 
pion candidates \cite{kaonid}. 
$K_S$ candidates are reconstructed in the $K_S \to \pi^+\pi^-$ decay
mode; the  pion combination is required to have 
an  invariant mass 
within $30\,$MeV$/c^2$ 
of the nominal $K_S$ mass and 
a vertex displaced  from the IP. 
The mass of the $D^\pm$ meson candidate is required to be within $10\,$MeV$/c^2$
($2.4\sigma$) of the nominal $D^\pm$  mass. 
 We select $B$ meson candidates using the energy difference
   $\Delta E = E^*_{B} - E^*_{\rm  beam}$ and the
   beam-energy-constrained mass
   $M_{\rm bc} = \sqrt{\left(E^*_{\rm beam}/c^2\right)^2 - 
   \left(p^*_{B}/c\right)^2}$,
   where $E^{*}_{B} $, $E^{*}_{\rm  beam}$, and $p^{*}_{B}$ are
   the $B$ meson energy, the beam energy, and the $B$ meson momentum,
   respectively, in the CM system.
 
The $K_S$ decay vertex is fitted from  two pion tracks.  
The $D^+$ meson decay vertex is fitted from three charged tracks  
or from the $K_S$ and $\pi^+$  track.
The mass of the
$K^-\pi^+\pi^+$ or $K_S\pi^+$ combination is constrained to the $D^+$ meson mass
to obtain better  $M_{\rm bc}$ and $\Delta E$ resolutions.
The $B^0$ decay vertex is reconstructed from the two $D$ meson tracks and the
IP information. All remaining charged tracks are used to determine the
decay vertex of the tag-side $B$ meson.
A loose requirement on the quality of the vertex fit is
applied for both $B$ mesons.  The reconstruction of the $B_{\mathrm{tag}}$ vertex, 
vertex quality and flavor tagging are not required  for the branching fraction measurement.  

The flavor of the accompanying  $B$ meson is determined from 
its decay products.
Events are divided into six \mbox{$r$-bins}
according to the tagging quality $r$. The value of $r$ ranges from $0$ for
events with no flavor information to $1$ for unambiguous flavor assignment. 
Due to the imperfect flavor tagging, the distribution  $\mathcal P_{\rm sig} $ of 
 Eq. \eqref{eq:dP/dt} is modified to 
\begin{eqnarray}
\mathcal P_{\rm sig} &=& 
\frac{e^{-|\Delta
  t|/\tau}}{4\tau}
    \{ 1 - q \Delta w + q (1 - 2  w)
     \label{eq:dP/dt-wtag} \\
   && \quad   [ \mathcal S \sin (\Delta
m \Delta t) + \mathcal A \cos (\Delta m \Delta t)]  \} \,\textrm{,}\nonumber
\end{eqnarray}
where $w$ is the wrong tag fraction, and  $\Delta w$   
is the difference between the wrong tag fractions if  the  $B_{\rm tag}$ meson is 
a $\overline{B}{}^0$ or $B^0$.
The values of $ w$
and $\Delta  w$ for each of the six bins in the tagging quality parameter $r$ 
are determined 
separately using flavor specific $B$ meson decays \cite{TaggingNIM}. 

Continuum events are suppressed by forming a likelihood ratio from
$\cos\theta_B$, where $\theta_B$ is the polar angle between the $B$ meson
direction in the CM system and the beam axis, and a variable based on a
combination of  sixteen  modified Fox-Wolfram moments with 
the scalar sum of transverse momentum  \cite{KSFW}.  
Note that since the $B\overline{B}$ and continuum events have significantly
different distributions in the tagging quality variable $r$, the continuum
suppression cut varies for events in different $r$-bins.

After applying all of the event selection criteria, 16\% of the signal events have more
than one $B^0$ candidate. The $B^0$ with the smallest value of 
  $\left(  \Delta m _{D^+} / \sigma_{D^+} \right)^2 + \left( \Delta m _{D^-} / \sigma_{D^-} \right)^2$ 
  is selected as the best candidate,   where $\Delta m
_{D} = M_{K\pi\pi/K_S\pi}-m_D$ is the difference from the nominal $D$ meson mass
and $\sigma_{D^{\pm}}$ are the widths of the signal peak in the $M_{K\pi\pi/K_S\pi}$ 
mass distribution. 

\begin{figure}[htb]
\centerline{
\begin{tabular}{ccc}
\includegraphics[width=0.35\linewidth]{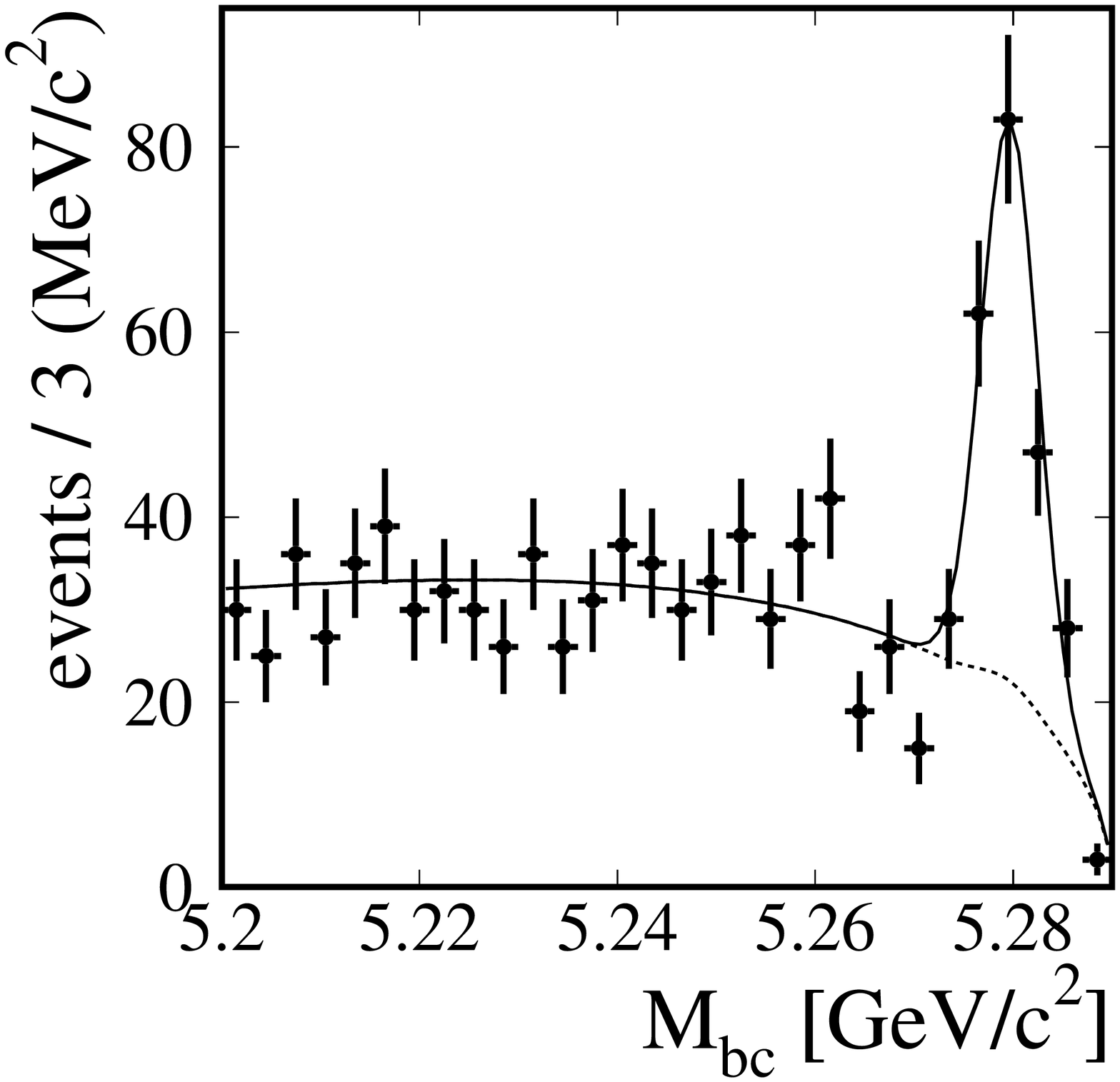} & \hspace{0.cm} &  
\includegraphics[width=0.35\linewidth]{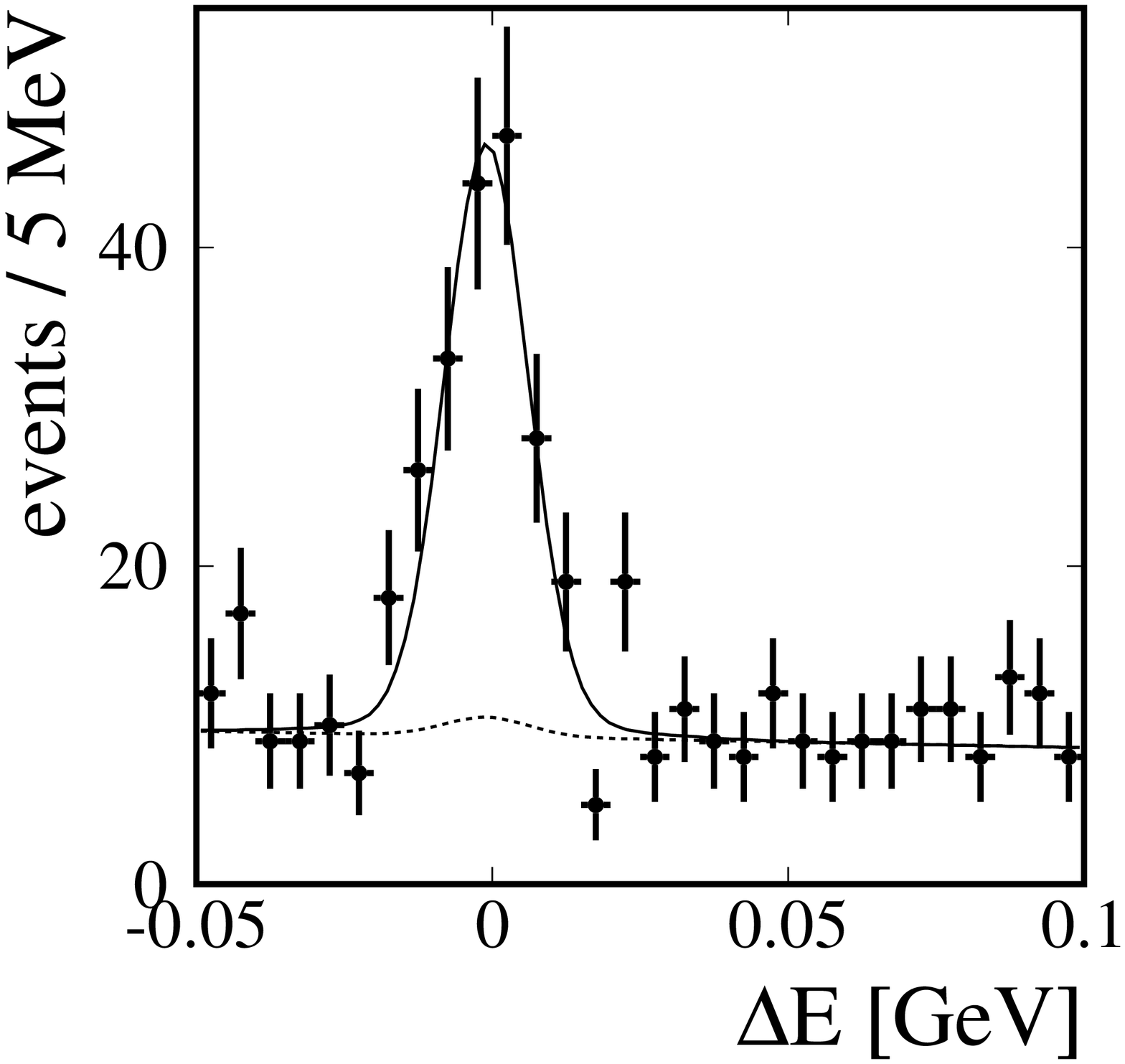} \\
(a) $M_{\rm bc}$, $|\Delta E| < 0.03 \,\mathrm{GeV}$ & & (b) $\Delta
E$, $M_{\rm bc} >  5.27\,\mathrm{GeV}/c^2$
\end{tabular}
}
\caption{Distributions for the reconstructed events in 
$M_{\rm bc}c^2$ (a) and  $\Delta E$ (b). The full (dashed) curves show
  the projections of the result of the 2D unbinned 
  maximum likelihood  fit for all (background) events.  
}
\label{fig:mbcde-data}
\end{figure}

The signal yield is obtained from an extended unbinned 2D maximum likelihood (ML) fit of the 
$M_{\rm bc}$ and $\Delta E$ distributions in the range 
$M_{\rm bc} > 5.20 \,\mathrm{GeV} / c^2$ and 
$-0.05 \,\mathrm{GeV} < \Delta E < 0.10 \,\mathrm{GeV} $.
A Gaussian function for the signal 
and an ARGUS \cite{argus} function for the background are 
used to describe the $M_{\rm
  bc}$ distribution.
For the  parameterization  of the
 $\Delta E$ distribution we used two Gaussians with the same mean value 
 to describe the signal and a linear   function to describe the background.
The fraction and the width of the wider Gaussian were fixed to the values obtained 
from Monte Carlo (MC) simulated signal decays \cite{MC}. 
Non-resonant $B^0 \to D^- \overline{K}{}^0 \pi^+ $ and $B^0 \to D^-
\overline{K}{}^{*0}(892) \pi^+ $ decays are  found to be a
possible source of background peaking in the  $M_{\rm bc}$ and $\Delta E$
distributions. The amount of this background was
estimated from the $D^+$ mass sidebands 
in data and subtracted from the signal. We estimate
the number of non-resonant decays in the signal region ($N_{\rm nr}$) 
to be  $2.0 \pm 1.8$  and $1.4 \pm 1.0$ for the $K\pi\pi$  and  $K_S\pi$ 
 channels, respectively. 
The fit yields 
$150 \pm 15$ events in the signal peak, where the error is statistical  only. 
The  $M_{\rm bc}$ and $\Delta E$ 
distributions of 
reconstructed events and the projection of the fit result
are shown in Fig.~\ref{fig:mbcde-data}. 
The signal yields from separate fits to the 
$K\pi\pi$ and $K_S\pi$ decay modes are given in Table~\ref{table:yield}. 
\begin{table}[htb]
\begin{center}
\caption{The product of $D$ branching fractions 
$\mathcal{B}_{D^+} \times \mathcal{B}_{D^-}$, 
the detection efficiency $\epsilon$, the number of
events in the signal peak  $N_\mathrm{peak}$ and
the expected amount of the combinatorial background $N_\mathrm{bcg}$ in
the $5.27 \,\mathrm{GeV} / c^2 < M_{\rm bc} < 5.29 \,\mathrm{GeV} / c^2 $ and 
$| \Delta E | < 0.03 \,\mathrm{GeV} $  region, as extrapolated from the fit.  
\label{table:yield}} 
\begin{tabular}{c||l@{$\pm$}lcr@{$\pm$}rr@{$\pm$}r}
\hline\hline
channel & \multicolumn{2}{c}{ $\mathcal{B}_{D^+} \times \mathcal{B}_{D^-}$ } &
$\epsilon$ [\%] & \multicolumn{2}{c}{$N_\mathrm{peak}$} &
\multicolumn{2}{c}{ $N_\mathrm{bcg}$}\\
\hline 
$K\pi\pi$ & $(0.904$&$0.065)$\% & 12.6  & $124.1 $&$ 13.6$ &$ 110.8 $&$ 2.6 $ \\
$K_S\pi$ & $(0.204$&$0.015)$\% & 12.1 & $ 25.7 $&$ 5.7 $ &  $ 13.8 $&$ 0.9 $ \\
\hline\hline
\end{tabular}
\end{center}
\end{table}

The combined branching fraction 
is calculated from the total number of reconstructed events
and the average reconstruction efficiency, and is found to be
$\mathcal{B}(B^0\to D^+D^-) = \left[1.97 \pm 0.20 \,\textrm{(stat)}\, \pm
    0.20\,\textrm{(syst)}\,\right] \,\times 10^{-4} $, 
which is consistent with previous measurements \cite{gobinda, babar2} 
and has better accuracy.
The uncertainty in the $D$ meson branching fractions results in a $5\%$ 
systematic error.
The error in the pion and kaon track reconstruction efficiency was estimated 
using partially reconstructed $D^*$ decays. The errors are added linearly for all six 
pion and kaon tracks, which yields a $6\%$ uncertainty. 
The difference in  PID efficiency for the simulated and real data 
 is approximately  $1\%$ per track,  which gives a $6\%$ uncertainty. 
Smaller contributions come from the uncertainty in the  
$K_S$ selection efficiency ($1\%$), the number of $B\overline{B}$ events ($1.3\%$) and the 
 number of non-resonant decays ($1.5\%$).
 The total systematic error of $10\%$  is
 obtained from the quadratic sum of 
these uncertainties.

Time-dependent $CP$ violation parameters  are determined by an unbinned
ML fit to the $\Delta t$ distribution 
of $219$ events in the signal region  
$5.27 \,\mathrm{GeV} / c^2 < M_{\rm bc} < 5.29 \,\mathrm{GeV} / c^2 $ and 
$| \Delta E | < 0.03 \,\mathrm{GeV} $.
The $\Delta t$ distribution for signal events $\mathcal P_{\rm sig} $ described by
Eq. \eqref{eq:dP/dt-wtag} is modified by the inclusion of the background
 contribution and resolution effects. The event-by-event likelihood is given by
\begin{equation}
\mathcal L_{\rm ev} =   f_{\mathrm{sig}} \, \mathcal P_{\mathrm{sig}} \otimes
  \,\mathcal R  \, +\, 
  f_{\mathrm{nr}} \, \mathcal P_{\mathrm{nr}}
 \otimes
  \,\mathcal R  \, +\,  
f_{\mathrm{bcg}} \, \mathcal P_{\mathrm{bcg}} \otimes \, \mathcal R_{\mathrm{bcg}} \,\textrm{.} \label{eq:deltat} 
\end{equation}
Subscripts sig,   nr and  bcg 
refer to signal, non-resonant and combinatorial background 
components, respectively. 
The fractions $f_{\mathrm{i}} = f_{\mathrm{i}}(M_{\rm  bc},
\Delta E, r)$ are determined on an event-by-event basis, 
$  f_{\mathrm{sig}} + f_{\mathrm{nr}} + f_{\mathrm{bcg}} = 1 $.
The function $ \mathcal R$ describes
the detector resolution of the $\Delta t$ measurement. It takes into
account the error in the determination of both $B$ meson vertices as well
as  an additional kinematic smearing due to the momentum
of the $B$ meson in the CM system and the smearing of
the tag-side vertex due to the tracks originating from the secondary vertices. 
An additional wide Gaussian component  with  $\sigma \approx 20\,$ps
  is added to describe
a small fraction  of events (about $1\,\%$) with poorly reconstructed vertices.
A more detailed description  of the resolution function parameterization  
 can be found in Ref.~\cite{vertexres}. Resolution parameters for the $B_{CP}$ 
meson vertex are determined from a fit to the $\Delta t$ distribution of
kinematically similar $B^0 \to D_s^+ D^-$ decays. 

The fraction of the non-resonant decays $f_{\mathrm{nr}}$ is assumed to be proportional
to the signal fraction, \mbox{$f_{\mathrm{nr}} = a\, f_{\mathrm{sig}}$}, where $a = N_{\rm nr}/(N_{\rm peak} -N_{\rm nr})$  
and  $a_{K\pi\pi}=0.016$,  $a_{K_S\pi}=0.059$.
The $\Delta t$ distribution of the 
non-resonant $B^0 \to D^- \overline{K}{}^0 \pi^+ $ or  $B^0 \to D^- \overline{K}{}^{*0}(892) \pi^+ $
background   is  described  by 
the $\Delta t$ distribution for
signal  with the parameters $\mathcal S$ and $\mathcal A$ set  to zero.
We include the effect of possible $CP$ asymmetry 
of these modes in the systematic error.
About half of the combinatorial background events come
from  $B\overline{B}$ decays ($b\to c$ transition), which have an
exponential decay $\Delta t$ distribution. 
The other half are continuum $e^+e^- \to q\bar{q}$ events, for which the
 $\Delta t$ distribution contains a $\delta $-function component. 
Therefore, the $\Delta t$ distribution of the background  is described by
\begin{equation}
\mathcal P_{\rm bcg}  =  \frac{1}{2} \left[ 
 (1 - f_\delta) \, \frac{e^{-|\Delta
    t|/\tau_\mathrm{bcg}}}{2\tau_\mathrm{bcg}} + f_\delta \, \delta(\Delta t) \right]
\, \textrm{.}   \label{eq:bcg}  
\end{equation}
The background resolution function $\mathcal R_{\rm bcg}$ 
is taken to be a Gaussian.  
Parameters  describing the background distribution  are obtained
from a fit to the  $\Delta t$ distribution of the data sideband, $M_{\rm bc} 
< 5.27\,\mathrm{GeV}/c^2$ and $\Delta E > 0.06 \,\mathrm{GeV}$.

\begin{figure}[!h]
\centerline{
\begin{tabular}{ccc}
\includegraphics[width=0.35\linewidth]{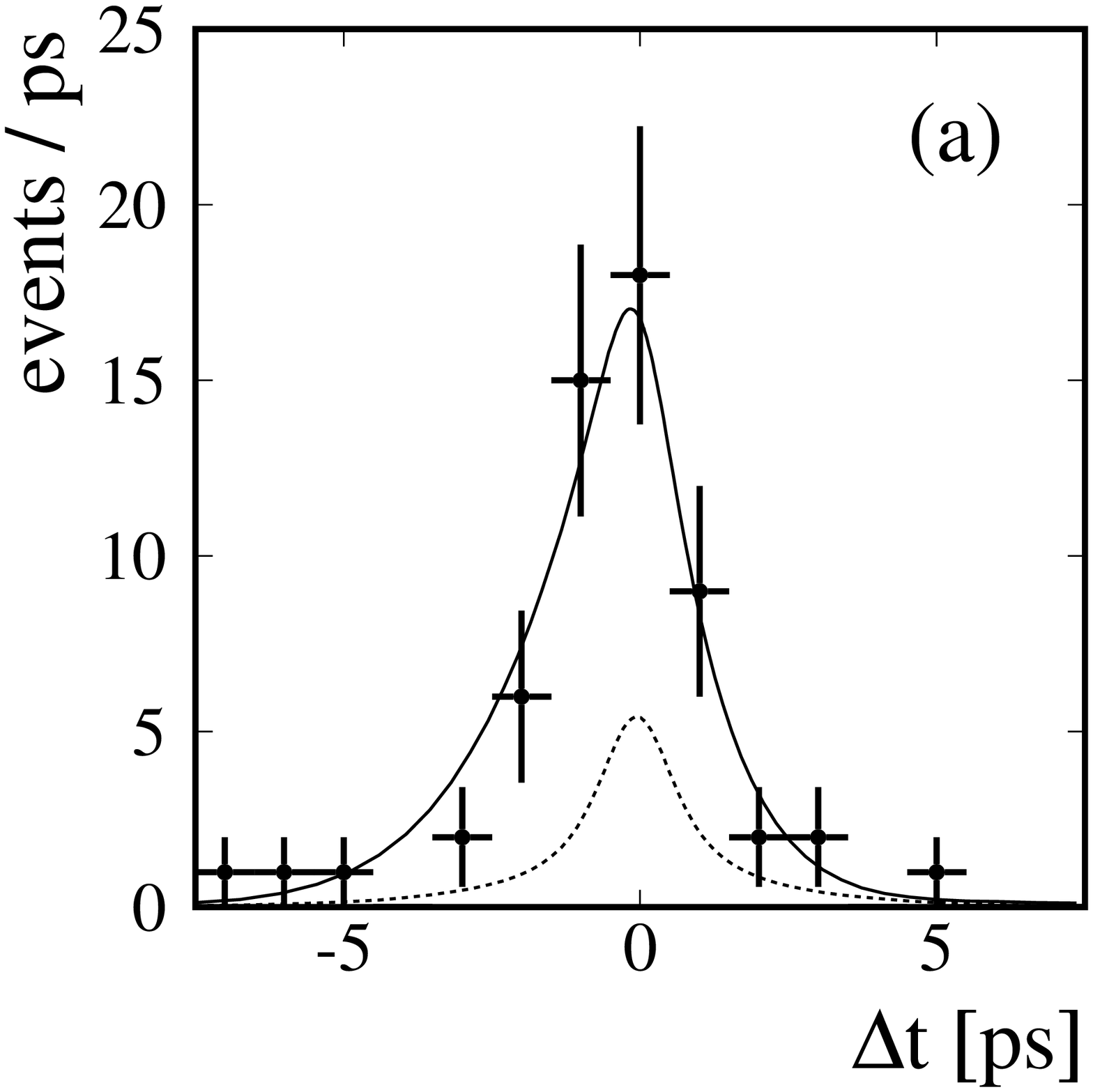} & \hspace{0.cm} &
\includegraphics[width=0.35\linewidth]{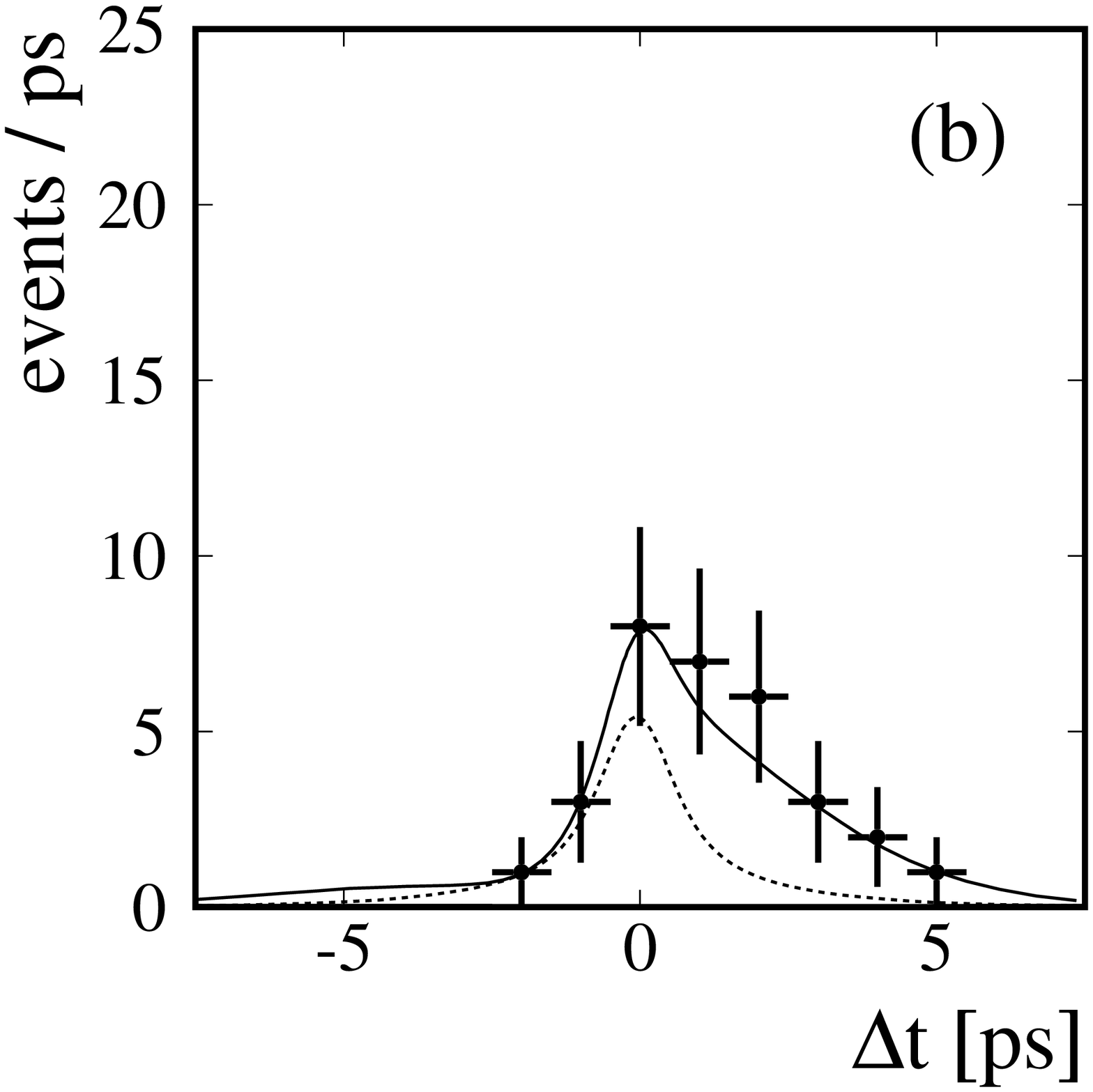}  
\end{tabular}
}
\caption{The $\Delta t $ distribution for 
events with good tagging information ($r > 0.5$)
when the tag-side $B$-meson is  
reconstructed as $B^0$ (a) or  $\overline{B}{}^0$ (b).  
The full and dashed curves show the projection of the
fit result and background contribution, respectively. 
} 
\label{fig:deltat}
\end{figure}

 From an unbinned  fit to the measured $\Delta t$ 
 distribution described by Eq. \eqref{eq:deltat}, 
we  obtain the $CP$ violation parameters for  $B^0 \to D^+ D^-$,  
\begin{eqnarray}
\mathcal S &=& -1.13 \, \pm 0.37 \,\pm 0.09 \,\textrm{ and }  \nonumber \\
\mathcal A &=& +0.91 \, \pm 0.23 \,\pm 0.06 \,\textrm{,}  
\end{eqnarray}
where the first error is statistical and the second is systematic.
The $\Delta t$ distributions are shown in Fig.~\ref{fig:deltat}. 
The main contributions to the systematic error are 
fit bias ($0.06$ for $\mathcal S$ and  $0.02$ for $\mathcal A$), 
uncertainties  in the resolution function ($0.04$ for $\mathcal S$ and 
$0.03$ for $\mathcal A$) and signal fraction ($0.035$ for $\mathcal S$ and
 $0.015$ for $\mathcal A$).  
 Other uncertainties come from
 the errors on the parameters $\tau$ and $\Delta m$ ($0.023$ for $\mathcal S$ and $0.007$ for $\mathcal A$), 
wrong tag fractions ($0.017$ for $\mathcal S$ and $0.014$ for $\mathcal A$),
description of background $\Delta t$ distribution ($0.01$ for $\mathcal S$ and $\mathcal A$), 
fraction and possible $CP$ asymmetry of the non-resonant background ($0.02$ for $\mathcal S$ and $0.03$ for  $\mathcal A$),
 the effect of tag-side interference \cite{tag-interference}  
($0.01$ for $\mathcal S$ and $0.03$ for $\mathcal A$) 
and requirements on 
the vertex quality and the fitting range (less than $0.01$ for $\mathcal S$ and $0.01$ for $\mathcal A$).

To test the consistency of the fitting procedure, the same analysis 
was applied to the  $B^0 \to D_s^+ D^-$  control sample. 
Since there is only one decay amplitude at the tree level and the leading
penguin contributions have the same CKM structure as the tree contribution,
no $CPV$ is expected for this decay.  
The result is consistent with no $CPV$, 
$ \mathcal S = -0.064 \pm 0.094  $ 
and  $\mathcal A = 0.091 \pm 0.060 $, where the  error is statistical only.
We also fit the background sample 
($M_{\rm bc} < 5.27\,\mathrm{GeV}/c^2$ and $\Delta E > 0.06 \,\mathrm{GeV}$) 
for a possible $CP$ asymmetry and find none: $\mathcal A=-0.01\pm 0.06$ and
$\mathcal S=0.03\pm 0.10$.
 In addition, 
a time-integrated fit for the parameter $\mathcal A$ was performed to validate the result
in $B^0\to D^+D^-$ decays.
The signal yield was determined separately for events tagged as  
$B_\mathrm{tag} = B^0 $ and $ B_\mathrm{tag} = \overline{B}{}^0 $  for  each
of the six $r$-bins. The fit  yields  
$\mathcal A =0.86 \pm 0.32$, which is consistent with the time-dependent result.

We use the Feldman-Cousins frequentist approach \cite{ref:feldmancousins} to 
determine the statistical significance  of our measurement. 
In  order to form confidence intervals, we use the $\mathcal A$ and $\mathcal S$  
distributions of the results of fits to the MC pseudo-experiments for various 
input values of $\mathcal A$ and $\mathcal S$ in a  similar way as described in 
Ref.~\cite{ref:pipiPRD93}. 
The systematic errors and possibility of tails that are  
wider than Gaussian tails are taken into account. 
The case of no $CPV$, $\mathcal S=\mathcal A=0$, is ruled out at the 
$4.1 \sigma$ confidence level (CL).
 The case of no direct $CPV$, $\mathcal A =0$, is excluded at more than 
 $3.2 \sigma$  CL
 for any value of the parameter $\mathcal S$.

In summary, we  
measure the branching
fraction   for  $B^0 \to D^+ D^-$  decays 
  to be $  \left(1.97 \pm 0.20 \, \pm  0.20\,\right) \,\times 10^{-4} $, 
  superseding our previous measurement \cite{gobinda}. 
We obtain values for the $CP$ parameters $\mathcal S =-1.13 \, \pm 0.37 \,\pm 0.09 $ and 
$\mathcal A = 0.91 \, \pm 0.23 \,\pm 0.06 $
 and rule out the $CP$-conserving case, $\mathcal S =\mathcal A =0$, at the 
 $4.1\,\sigma$  confidence level.  
 The value of $\mathcal S$ is consistent with 
 measurements of  $b \to c \bar{c} s$  modes \cite{PDG}.   
In addition, we observe evidence for direct $CP$ violation at the $3.2\,\sigma$ confidence level. Some extensions of the SM 
 predict  large contributions to the $CP$ violating phases in $b\to c\bar{c}d$ decays 
 that are 
 consistent with our result  \cite{ref:beyondSM}.
 Our measurement   
 differs from a previous measurement by the BaBar collaboration \cite{babar} 
  by about $2.2\,\sigma$. 
   
We thank the KEKB group for excellent operation of the
accelerator, the KEK cryogenics group for efficient solenoid
operations, and the KEK computer group and
the NII for valuable computing and Super-SINET network
support.  We acknowledge support from MEXT and JSPS (Japan);
ARC and DEST (Australia); NSFC and KIP of CAS (China); 
DST (India); MOEHRD, KOSEF and KRF (Korea); 
KBN (Poland); MIST (Russia); ARRS (Slovenia); SNSF (Switzerland); 
NSC and MOE (Taiwan); and DOE (USA).


\begin{thebibliography}{99}
\bibitem{beta} Another naming convention, 
$\beta = \phi_1$, is also used in the literature.
\bibitem{CC}
The inclusion of the charge conjugate mode decay is implied throughout this paper.
\end{thebibliography}
\end{document}